\documentclass[12pt]{article}
\usepackage{amsmath}
\usepackage{cite}
\usepackage{bbold}
\usepackage{amssymb}
\usepackage{epsfig,graphicx}
\usepackage{graphicx}
\usepackage{capt-of}
\usepackage{float}
\usepackage{lipsum}
\usepackage{subcaption}
\usepackage{tikz}
\usepackage{float}
\usepackage{diagbox}
\usepackage{multirow}

\def\be{\begin{equation}}
\def\ee{\end{equation}}
\def\bea{\begin{eqnarray}}
\def\eea{\end{eqnarray}}

\begin{document}
\begin{center}
{\large \bf SYK model with an extra diagonal perturbation: \\
phase transition in the eigenvalue spectrum}\vskip 0.5cm

{\large \bf Shuang Wu}\footnote{scholes89@sjtu.edu.cn}
\\[0.1cm]
{\it Shanghai Center for Complex Physics, School of Physics and Astronomy,
Shanghai Jiao Tong University, Shanghai 200240, China}
\\[0.1cm]
\end{center}

\vskip 0.5cm
\centerline{\large \bf Abstract}
\vskip 0.2cm  
We study the SYK model with an extra constant source,  \.i.e. a constant matrix or equivalently a diagonal matrix with only one non-zero entry  $\lambda_1$.
By using methods from analytic combinatorics \cite{Flajolet2}, we find exact expressions for the moments of this model. We further prove that the spectrum of this model can have a gap when $\lambda_1>\lambda^c_1$, thus exhibiting a phase transition in $\lambda_1$. In this case, a single isolated eigenvalue splits off from SYK's eigenvalues distribution. We located this single eigenvalue by analyzing the singular behavior of a supercritical functional composition scheme. In certain limit our results recover the ones of random matrices with non-zero mean entries. 

\section{Introduction}
Considering the SYK model plus a diagonal matrix with only one non-zero entry $\lambda_1$:
\begin{align}
H:&=H_{\text{SYK}}+\mathbb{\Lambda}\nonumber\\
&=i^{p(p-1)/2}\sum_{1\leq i_1<i_2<\dots<i_p\leq N} J_{i_1,i_2,\dots,i_p} \psi_{i_1}\psi_{i_2}\dots\psi_{i_p}+\begin{bmatrix}\lambda_1 &  & 0\\   &  \ddots & \\ 0 & & 0\end{bmatrix}\label{hamiltonian}
\end{align}
the first part above is just the usual SYK\cite{Sachdev1,Sachdev2,Kitaev} Hamiltonian with $p$ $\psi_{j}$  interacting Majorana fermions among $N$
Majorana fermions satisfying $\lbrace\psi_{i},\psi_{j}\rbrace=2\delta_{i,j}$. The factor $i$ in front of \eqref{hamiltonian} is introduced to keep $H_{\text{SYK}}$ Hermitian.
$J_{i_1,i_2,\dots,i_p}$ are random coupling satisfying
\begin{align}
\langle J_{i_1,i_2,\dots,i_p}^2\rangle_{J}=\frac{1}{\binom{N}{p}}
\label{Jsclae}
\end{align}
where $\langle \cdots\rangle_{J}$ represents the ensemble average over the random couplings $J$ \footnote{Note that the variance \eqref{Jsclae} is not a common choice. In the double scaled limit \eqref{double-scaled}, it is different from the usual SYK conventions only by a constant factor\cite{Berkooz}. This is a normalization factor which corresponds to the number of terms under the sum over the index sets in \eqref{hamiltonian} and is chosen to keep the spectrum of order $1$.}. We choose this scaling to simplify the calculation, with \eqref{Jsclae}, we normalize the trace of SYK Hamiltonian $ \langle \text{tr} H_{\text{SYK}}^2 \rangle_J  =1$\footnote{trace operator in this paper is normalized as $\text{tr} \mathbb{1}=1$.}
. Therefore, moments calculation can be reduced to a combinatorial counting problem \cite{Berkooz}.
The last term $\mathbb{\Lambda}$ in \eqref{hamiltonian} is a $2^{\lfloor N/2\rfloor}\times 2^{\lfloor N/2\rfloor}$ diagonal matrix with only one non-zero element $\lambda_1$. Since the SYK Hamiltonian is a large sparse matrix, \eqref{hamiltonian} can also be treated as a matrix. We note that \eqref{hamiltonian} can be understood as a special case of an SYK-model deformed by a random Hamiltonian diagonal in Fock space \cite{Altland}. The model in \cite{Altland} is non-ergodic, namely 
its spectral statistics is of Wigner-Dyson type, whereas its wave functions are non-uniformly distributed over Fock space. It serves as a toy model to study the non-ergodic extendedness  in quantum chaotic systems. The perturbation in this model is introduced by a  diagonal term \footnote{in \cite{Altland}, they call it Fock-space diagonal operator $\tilde{H_v}=\gamma\sum_{n}v_n\vert n\rangle\langle n\vert$.}, which
is a sum over projectors onto the occupation number eigenstates
$\vert n\rangle=\vert n_1,n_2,\cdots,n_k,\cdots\rangle$, with $n_i=0,1$. Each state attached to a coefficient $v_n$ as a random variable with respect to a box distribution. In their case, the spectral statistics are of Wigner-Dyson type.
Now in our case, $\mathbb{\Lambda}$ can be understood as a special projection of one eigenstate, which is the ground state $\vert 1,0,0,\cdots\rangle$ coupled with a fixed eigenvalue $\lambda_1$. The crucial difference between \eqref{hamiltonian} and the model in \cite{Altland} is that we fixed the value of $\lambda_1$; they choose a random variable $v_n$ to attach the occupied state $\vert n\rangle$. Thus, there is a fundamental difference between the spectral densities of these two model. Only a fixed coefficient can bring the largest eigenvalue separation into the spectrum. The reason is that there is a phase transition in $\lambda_1$, only when $\lambda_1$ passes its critical value, a single eigenvalue splits off from the main spectrum of \eqref{hamiltonian}. This is why the model in \cite{Altland} has no such separation in its spectral density.

There are two motivations of studying this model. The first one is to find 
an SYK-like model which has similar spectral properties of random matrix with non-zero mean entries\cite{edwards,jones}. Like random matrix, a small rank perturbation of SYK can affect significatively the limiting properties of the spectrum as the system size goes to infinity. Depending on the values of $\lambda_1$ we put in, we can have a separation of the largest eigenvalue, thus there is a phase transition in $\lambda_1$ just like the random matrix case. The other
reason for addressing this rather simple perturbatively model arises
from the deep connection between the
phase transitions in physics and the asymptotic analysis of a critical composition scheme of combinatorial structures (\.i.e. when a class of combinatorial structures can be decomposed into more basic building blocks, its generating function is then a functional composition of those of the building blocks. It is critical when both its and the building blocks generators are simultaneously singular). See \cite{Banderier} for more recent discussion about this connection.

In this work, we first map the ensemble averaging moments calculation of \eqref{hamiltonian} into a counting problem of complex combinatorial structures. Furthermore, the class of these structures can be obtained by combining operators of more well-studied classes of substructures. Any specification of a constructible class  translates directly into its generating function equations. Then, we can enumerate this decomposable class via its generating function. In our case, the specific class is constructed by a pointing operator of a labeled cyclic class of certain rooted substuctures. In section \ref{sec:1} we will give a brief introduction about how to construct a decomposable class of combinatorial structures by using specific operations which will be used in our work. We refer \cite{Flajolet2} for detailed definitions. 

In the following, we are interested in studying the distribution of eigenvalues of \eqref{hamiltonian}, in the doubled scaled limit \cite{Garcia1,Berkooz,Erdos}
\begin{equation}
N \rightarrow \infty, \quad \lambda=\frac{2 p^{2}}{N}=\text { fixed. }\label{double-scaled}
\end{equation}
First of all,  define $\Psi_{\alpha}=i^{p(p-1)/2}\prod_{i=1}^p\psi_{i_i}$ as the product of $p$ Majorana fermions, $\alpha$ represents an index set
as $\alpha=\left\lbrace i_{1},i_{2}\cdots i_{p}\right\rbrace $ which satisfy
\begin{align}
\Psi_{\alpha}^{2}=\mathbb{1}, \quad \Psi_{\alpha} \Psi_{\beta}=(-1)^{c_{\alpha \beta}+p} \Psi_{\beta} \Psi_{\alpha}\label{gammarel}
\end{align}
where $c_{\alpha \beta}=\vert \alpha\cap\beta\vert$ is the  number of common indices in the sets $\alpha$ and $\beta$. In the doubled scaled  limit, $c_{\alpha \beta}$ follows a Poisson distribution with mean $p^2/N$\cite{Berkooz}. 
In the following we assume $p$ is always even. In addition, we introduce a new variable $q$ as a factor to count the exchange between any two $\Psi$ in this double scaling limit \eqref{double-scaled}
\begin{align}
&\Psi_{\alpha} \Psi_{\beta}=q \Psi_{\beta} \Psi_{\alpha}\quad\forall\alpha ,\beta\nonumber\\
\text{with}\quad &q:=\sum_{c_{\alpha \beta}=0}^{\infty}\frac{(p^2/N)^{c_{\alpha \beta}}}{c_{\alpha \beta}!}(-1)^{c_{\alpha \beta}}\exp(-p^2/N)=\exp(-\lambda)
\label{q}
\end{align}
therefore $q\in[0,1]$. We also assume that no three sets of fermions can have common elements at the same time $\vert\alpha\cap\beta\cap\gamma\vert=0$. This is equivalent to assuming that no three chords in a diagram intersect at the same point.
Finally, we should point out that though our results are computed in the
limit \eqref{double-scaled}, they are  also have a very good agreement with fixed finite $p$ cases (see similar situation in \cite{Garcia1, Garcia2}). The trick is with finite $p$, one has to change the notation of $q$ in \eqref{q} to
\begin{align}
\tilde{q}(p)=\sum_{c=0}^{p}(-1)^{c}\binom{p}{c}\binom{N-p}{p-c}
\label{qbis}
\end{align} 
\subsection{Results}
\label{subsec:2}
The main results in the paper are the following.

Define $m_p$ as the ensemble averaging moments of the \eqref{hamiltonian} extracted by ensemble averaging moments of the SYK Hamiltonian
$$
m_p=2^{\lfloor N/2\rfloor}\left(\langle \text{tr} H^p \rangle_J -\langle \text{tr} H_{\text{SYK}}^{2\lfloor  p/2\rfloor} \rangle_J \right)
$$
\begin{align}
m_p =\sum_{j=0}^{\lfloor\frac{p-1}{2}\rfloor}\lambda_1^{p-2j}\frac{p}{p-2j}\sum_{\substack{k_1,k_2,\ldots,k_j \geq 0 \\ k_1+2k_2+\cdots + jk_j = j}} \binom{p-2j}{k_1,k_2,\ldots,k_j,p-2j-\sum_{l=1}^j k_l} \prod_{i=1}^j \text{RT}(i,q)^{k_i}
\label{moment1}
\end{align}
where $\lfloor\frac{p-1}{2}\rfloor$ denotes the integerpart of $\frac{p-1}{2}$. $\text{RT}(i,q) = \langle \text{tr} H_{\text{SYK}}^{2i} \rangle_J$ is the Riordan-Touchard formula, which counts the ensemble averaging $2i$-th normalized moment of SYK in the double scaled limit\cite{Garcia1,Berkooz,Erdos}. We notice that when $q=0$, $\text{RT}(i,0)$ is the $i$-th Catalan number, \eqref{moment1} can be simplified dramatically, it recovers the moments of the eigenvalue density of a random matrix with a finite mean \cite{jones}  (see appendix for the proof).
By rearranging the second sum in \eqref{moment1}, we get
\begin{align}
m_p  =\sum_{j=0}^{\lfloor \frac{p-1}{2}\rfloor}\lambda_1^{p-2j}\frac{p}{p-2j}\sum_{\substack{l=1 \\ k_1+k_2+\cdots + k_l = j}}^j  \binom{p-2j}{l} \prod_{i=1}^l \text{RT}(k_i,q)
\label{moment2}
\end{align}
the second sum in \eqref{moment2}, is a sum over all compositions of integer $j$ into $l$ parts. It should be noted that the expressions above are not special, other non-commutative models also have these kinds of structures (for ex, Hofstatder model\cite{Ouvry}).

We prove the generating function to be
\begin{align}
\sum_{p\geq 1}m_p z^p=z\frac{d}{dz}\log\frac{1}{1-\lambda_1zR(z,q)}
\label{generationgF}
\end{align}
where $R(z,q)$ is defined as the generation function of $\text{RT}(i,q)$
\begin{align}
R(z,q):=\sum_{i\geq 0} \text{RT}(i,q) z^{2i}=\frac{\sqrt{1-q}}{z}\sum_{n\geq  0}(-1)^n q^{n(n+1)/2}\left(\frac{1-\sqrt{1-(4z^2/(1-q))}}{2z/\sqrt{1-q}}\right)^{2n+1}
\label{gf_SYK}
\end{align}
$R(z,q)$ is first given in \cite{Cappelli}, which is closely related to the Stieltjes transform of the measure of continuous q-Hermite polynomial\cite{Bustoz, Ismail}.
\eqref{generationgF} is equivalent to
\begin{align}
\sum_{p\geq 1}m_p z^p=\left(1+\frac{z R^{\prime}(z,q)}{R(z,q)}\right)\frac{\lambda_1 zR(z,q)}{1-\lambda_1 zR(z,q)}
\label{generationgF2}
\end{align}
where $R^{\prime}(z,q)=\frac{dR(z,q)}{dz}$. This is because every decomposable structure admits an equivalent standard specification \cite{Flajolet3}. See the proof of the equivalence between \eqref{generationgF} and \eqref{generationgF2}
in section \ref{subsec3.1}.
We note that the generating function in \eqref{generationgF2} can be  approximated by
\begin{align}
\int_{-\infty}^{+\infty}\rho_{\texttt{extra}}(E)\frac{1}{1-z E}\text{d}E
\label{densityGF}
\end{align}
where $\rho_{\texttt{extra}}(E)$ is the approximated density of states corresponding to the moments $m_p$:
\begin{align}
\rho_{\texttt{extra}}(E)=\left(\frac{E}{\lambda_1R(1/E,q)}\right)^{\prime}\delta\left(\frac{E}{\lambda_1R(1/E,q)}-1\right).
\label{density}
\end{align}
One can show that \eqref{density} is obtained from \eqref{densityGF} because of a simple identity
$$
\int_{-\infty}^{+\infty}\frac{\delta(s-1)}{1-z s}\text{d}s=\frac{1}{1-z}.
$$
Set $z=\lambda_1 z^{\prime} R(z^{\prime},q) $, then $z^{\prime}\frac{d z}{dz^{\prime}}=\lambda_1 z^{\prime}R(z^{\prime},q)(1+\frac{z R(z^{\prime},q)^{\prime}}{R(z^{\prime},q)})$.

Furthermore, the Kronecker delta function in \eqref{density} gives us the location of the separated eigenvalue, which is the solution of the secular equation
\begin{align}
\frac{E}{\lambda_1R(1/E,q)}-1=0.
\label{ageq}
\end{align}
We notice that \eqref{ageq} can only have a solution with $E>\frac{2}{\sqrt{1-q}}$ when $\lambda_1>\lambda^c_1$ which will be identified in \ref{subsec:3.1}.
For example, when $q=0$, $R(z,0)=\frac{1-\sqrt{1-4z^2}}{2z^2}$, which is the generating function for the Catalan numbers. Therefore, \eqref{ageq} in this case can have a solution $\frac{1+\lambda_1^2}{\lambda_1}$ only when $\lambda_1>1$. This is exactly the same as the random matrix theory with shifted mean Gaussian entries \cite{edwards,jones}.

To conclude, the ensemble averaging density of \eqref{hamiltonian} in the  double scaled limit \eqref{double-scaled} is
\begin{align}
\rho(E)=\left\{\begin{array}{ll}\rho_{0}(E)+2^{-\lfloor N/2\rfloor} \delta\left(E-E_{\text{split}}\right) \quad & \lambda_1>\lambda^c_1 \\ \rho_{0}(E) \quad & \lambda_1<\lambda^c_1\end{array}\right.
\label{densityF}
\end{align}
where $E_{\text{split}}$ is the solution of \eqref{ageq} and 
$\rho_{0}(E)$ denotes the ensemble averaging density of SYK model in the  double scaled limit \eqref{double-scaled}, which is given by the weight function of Q-Hermite polynomial \cite{Garcia1}
\begin{align}
\rho_{0}(E):=\rho_{QH}(E)=c \sqrt{1-\left(\frac{E}{2/\sqrt{1-q}}\right)^{2}} \prod_{k=1}^{\infty}\left[1-4 \left(\frac{E}{2/\sqrt{1-q}}\right)^{2}\left(\frac{1}{2+q^{k}+q^{-k}}\right)\right]
\label{densitySYK}
\end{align}
where $c$ is a normalization constant such that $\int \rho_{0}(E)dE=1$.
Once we get $E_\text{split}$ the location of the separated eigenvalue, the asymptotic expression of the moments $m_p$ can be simplified significantly
\begin{align}
m_p\simeq \lambda_1^p\sum_{j=0}^{\lfloor\frac{p-1}{2}\rfloor}\binom{p}{j}\left(\frac{E_\text{split}}{\lambda_1}-1\right)^j
\label{moment4}
\end{align}
Both \eqref{moment1} and \eqref{moment4} are in closed agreement  with the numerical results with fixed $p$, the comparisons between analytical and numerical results are given in section  \ref{sec:4}.
\subsection{Plan of the paper}
In section \ref{sec:1}, we introduce briefly the basic combinatorial languages to be used throughout this paper. Section \ref{sec:2} is devoted to explaining in detail the mapping between moments calculation of $m_p$ and the enumeration of certain combinatorial structures identified below. We collect results of $m_p$  from analytic combinatorics. Section \ref{sec:3} contains the singularity analysis of the generating function of combinatorial structures. We further demonstrate how to locate the phase transition and the gap in the spectrum of \eqref{hamiltonian} by using singularity expansions. Section \ref{sec:4} presents the comparison between our analytic expressions and the numerical results obtained by calculating the eigenvalues of \eqref{hamiltonian} with Mathematica.
Finally, in section \ref{sec:5}, we give a brief conclusion and discuss the directions for further research.  
\section{Combinatorial structures and construction operators}
\label{sec:1}
First of all, a class of  combinatorial structures is a set of objects with different sizes. The size of an object is the number of its nodes (components). In addition, the number of objects of each size is finite. The set is described by finite rules. In order to enumerate these objects with respect to their properties, these rules admit a direct translation as operations over generating functions of corresponding structures. On top of this, many combinatorial structures can  be built by more elementary building blocks; this translate into generating function means that a decomposable structure has a functional composition scheme for its generation function. In  the following we present some basic concepts about how to construct a combinatorial structure by putting  more basic substructures together.
\subsection{Labelling object}
\label{subsec2.1}
A labeled object means each of its nodes carries, whether with a distinctive color, or equivalently
an integer label, in such a way that all the labels occurring in an object are distinct. So a labeled class is a collection of labeled objects with different sizes. 
An object is rooted if a particular node is specified, this node is known as the root. Therefore for a rooted structure $\mathcal{A}$, its generating function is
$$
\mathcal{A}=\mathcal{Z}\operatorname{\times}\mathcal{B}\quad\Longrightarrow \quad A(z)=zB(z)
$$
where $\mathcal{Z}$ represents the root in $\mathcal{A}$, it is then attached to substructure $\mathcal{B}$. $A(z)=\sum_{k}a_k z^k$  and $B(z)=\sum_{k}b_k z^k$ are two generation functions enumerate structurres $\mathcal{A}$ and $\mathcal{B}$ respectively (\.i.e. the coefficients $a_k$ and $b_k$ are the numbers of $\mathcal{A}$-structures and of $\mathcal{B}$-structures of size $k$ respectively). 
\subsection{Construction operators}
\label{subsec2.2}
Let's consider two basic operations: cartesian product $\operatorname{\times}$ and sequence $\operatorname{SEQ}$,
\begin{align}
\mathcal{A}=\mathcal{B}\operatorname{\times}\mathcal{C}\quad\Longrightarrow \quad A(z)=B(z)C(z)
\label{op_product}
\end{align}
this means the cartesian product $\mathcal{A}$ consists in forming all pairs with a first component in $\mathcal{B}$ and a second
component in $\mathcal{C}$.
\begin{align}
\mathcal{A}=\operatorname{SEQ}(\mathcal{B})\quad\Longrightarrow \quad A(z)=\frac{1}{1-B(z)}
\label{op_seq}
\end{align}
$\operatorname{SEQ}(\mathcal{B})$ builds a sequence of component from $\mathcal{B}$.
Next, consider a labeled cycle structure $\operatorname{CYC}(\mathcal{B}):=\text{SEQ}(\mathcal{B})/\textbf{S}$, which is a set of cycles of labeled nodes; each node is attached to elementary building blocks $\mathcal{B}$ up to the equivalence relation $\textbf{S}$ (\.i.e., a cycle is invariant under the rotation), the corresponding generationg function of $\operatorname{CYC}(\mathcal{B})$  is
\begin{align}
\mathcal{A}=\operatorname{CYC}(\mathcal{B}) \quad \Longrightarrow \quad A(z)=\sum_{k=1}^{\infty} \frac{B(z)^{k}}{k} =\log \frac{1}{1-B(z)}
\label{op_cyc}
\end{align}
The final operation we need to construct our structure is the pointing operator $\Theta$. It consists in pointing (or marking) one of the components along all the ones that compose an objet of size $n$. $\Theta$ operator is equivalent to the differential operator in  combinatorial differential calculus \cite{Leroux}.
For a given structure $\mathcal{B}$ of  size $n$, the pointing of $\mathcal{B}$ means that one of its $n$ nodes is distinguished; hence, there are $n$ different ways of pointing this structure. Consider a structure {$\mathcal{A}$ which can be decomposed by a pointing operator of a substructure $\mathcal{B}$, its generationg function follows:
\begin{align}
\mathcal{A}=\Theta\mathcal{B}\quad\Longrightarrow \quad A(z)=z\frac{d}{dz}B(z)
\label{op_pointing}
\end{align}
see figure \ref{pointing} for the illustration. 
\begin{figure}[H]
\begin{center}
\includegraphics[scale=.5]{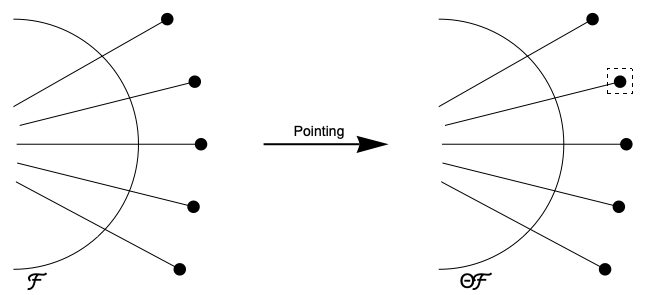}
\caption{An example of pointing operator of a given structure $\mathcal{F}$ with size $5$: one of the $5$ nodes is pointed by a dotted rectangle on the left side of the figure}
\label{pointing}
\end{center}
\end{figure}
\subsection{Multitudinous combinatorial structure}
\label{subsec2.3}
Now consider a complex structure $\mathcal{A}$ which is composed by a pointing operator of labeled cycles of a rooted substructure $\mathcal{B}$. Clearly,
\begin{align}
\mathcal{A}=\Theta\operatorname{CYC}\left(\mathcal{Z}\times\mathcal{B}\right)\quad\Longrightarrow \quad A(z)=z\frac{d}{dz}\log\frac{1}{1-zB(z)}
\label{op_complex}
\end{align}
this leads to the enumeration of the $\mathcal{A}$-structures of size $n$
\begin{align}
a_n&=n[z^n]\log \frac{1}{1-zB(z)}\nonumber\\
&=n\sum_{k_1+2k_2+3k_3+\cdots=n}\frac{\binom{k_1+k_2+k_3+\cdots}{k_1,k_2,k_3,\cdots}}{k_1+k_2+k_3+\cdots} b_1^{k_1}b_2^{k_2}b_3^{k_3}\cdots
\label{op_mix}
\end{align}
where $[z^n]$ denotes the coefficient of extraction of $z^n$ in series expansion.

Now we illustrate the previous definitions by enumerating a pointed cycle of binary trees. The building block here is the binary tree, \.i.e. ,  a rooted plane tree with each node having two descendants; see figure \ref{fig:binarytree} for some examples.
\begin{figure}[H]
\begin{center}
\includegraphics[scale=.7]{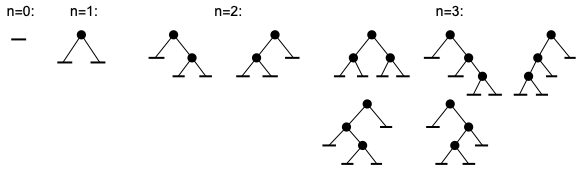}
\caption{Binary trees with size $n=0,1,2,3$ with
respective cardinalities $1,1,2,5$}
\label{fig:binarytree}
\end{center}
\end{figure}
The number of binary trees that have $n$ branching nodes is given by Catalan number $C_{n}$ and the corrresponding generationg function is
\begin{align}
B(z):&=\sum_{n \geq 0} C_{n} z^{n}\nonumber\\
&=\frac{1-\sqrt{1-4z}}{2z}
\label{CatGF}
\end{align} 
Now consider a pointed labeled cycle $\mathcal{A}$, which is a necklace with each labeled bead attached to a binary tree, \.i.e. the root of each tree connects directly to a labeled bead. Thus, according to \eqref{op_mix}, its generating function is 
\begin{align}
A(z):&=\sum_{n \geq 1} a_{n} z^{n}\nonumber\\
&=z\frac{d}{dz}\log\frac{1}{1-\frac{1-\sqrt{1-4z}}{2}}
\label{neckGF}
\end{align}
where $a_n$ is the number of such necklaces of size $n$. Therefore, by expanding \eqref{neckGF}, we get the enumeration of this complex structure.
Figure \eqref{fig:necklacetree} below shows an example of the enumeration of these necklaces with $n=3$. 
Construction of a labeled necklace with $3$ nodes in terms of binary trees: this necklace can either have 
\begin{itemize}
\item  $1$ labeled bead attached to a binary tree with $3$ nodes, therefore, it has 5 possibilities for the attachment (see the rightmost figure in Fig. \ref{fig:binarytree}).
\item
$2$ labeled beads, one of them attached to a binary tree with $2$ nodes (two possibilies for a binary tree of size $2$; see the middle figure in Fig. \ref{fig:binarytree}) and the other attached to a binary tree with $1$ node. In addition, since the necklace has two labeled beads, either of them can be attached to a binary tree of size $2$. So we have double chances for labelling. Therefore, we have $2(2\times 1)=4$ possibilities to compose this necklace.
\item $3$ labeled beads, each of them attached exactly to $1$ node. Since each bead is attached to the same structure, it is cyclic invariant under rotation. Therefore, we have exactly $1$ way to make this necklace.
\end{itemize}
To sum up, the total number of labeled necklaces with $3$ nodes in terms of binary trees is $5+4+1=10$, which equals exactly to $3[z^3]\log\frac{1}{1-zB(z)}$ with $B(z)$ in \eqref{CatGF}.
\begin{figure}[H]
\begin{center}
\includegraphics[scale=.5]{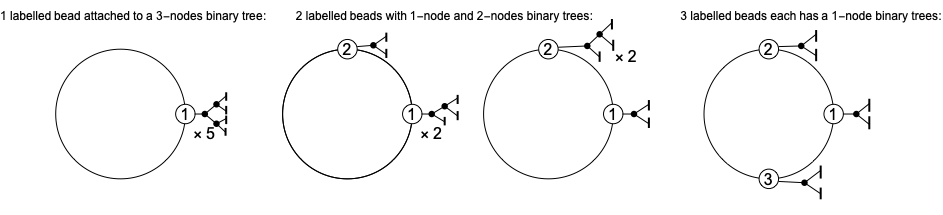}
\caption{All the possible labeled necklaces of binary trees of size $n=3$}
\label{fig:necklacetree}
\end{center}
\end{figure}

In the following section, we study
the mapping between the generationg function of certain combinatorial class and moments calculation of quantum mechanics model \eqref{hamiltonian} 
\begin{align}
\langle\text{tr}H^n\rangle_{J}=n[z^n]\log \frac{1}{1-zB(z)}
\label{mapping}
\end{align}
with certain generating function $B(z)$ which will be identified later.
Once we identify the specification of the combinatorial class, we can use singularity analysis of the corresponding generating function to locate the phase transition that happens in the eigenvalues distributions of \eqref{hamiltonian}.
\section{Moments calculation}
\label{sec:2}
In this section, we discuss how to study the eigenvalue distribution of \eqref{hamiltonian} by using analytic combinatorics \cite{Flajolet2} to construct a complex combinatorial structure. Our approach is first to map the ensemble averaging moments of \eqref{hamiltonian} into a counting problem of complex combinatorial structures; then we can get an analytic expression from the combinatorial counting. It is important to remember that our calculation is under the double scaled limit \eqref{double-scaled}. 
Our goal here is to calculate
\begin{align}
m_p:=2^{\lfloor N/2\rfloor}\left(\langle \text{tr} (H_{\text{SYK}}+\mathbb{\Lambda})^p \rangle_J-\langle \text{tr} H^{2\lfloor  p/2\rfloor}_{\text{SYK}}\rangle_J\right)
\label{moments3}
\end{align}
where $p$ can be either even or odd nonzero integers. Let's recall that  $\text{tr}$ in this paper is normalized such that $\text{tr}\mathbb{\Lambda}^i=\lambda_1^i/2^{\lfloor N/2\rfloor}$ and $\langle\text{tr} H_{\text{SYK}}^{2j}\rangle_J=\text{RT}(j,q)$ with $i$ and $j$ being positive integers. Thus, we need to multiply the difference in \eqref{moments3} by $2^{\lfloor N/2\rfloor}$ to get finite results in large $N$ limit.
Notice that the Hamiltonian of SYK can be treated as a sparse matrix; therefore we can not expand \eqref{moments3} in terms of binomial coefficients. We can map the calculation of \eqref{moments3} into a counting problem of cyclic composition, since the trace of a matrix is invariant under cyclic permutation.  

To begin with, the ensemble averaging operator $\langle\cdots\rangle_J$ only affects $H_{\text{SYK}}$; it forces all the $H_{\text{SYK}}$ in \eqref{moments3} to appear in pairs since every independent $J$ needs to appear exactly twice to obtain non-zero results after the ensemble averaging calculation. The pairing of all the
$H_{\text{SYK}}$ in \eqref{moments3} leads to a counting problem of perfect matching.  Consequently, the contribution from $H_{\text{SYK}}$ in \eqref{moments3} must be even. 
Thus one can represent $\langle\text{tr} H_{\text{SYK}}^k\rangle_J$ by enumerating chord diagrams. Each diagram is a circle with $k$ even nodes. Each node represents an $H_{\text{SYK}}$  insertion, then using chords to connect the nodes in pairs when two $H_{\text{SYK}}$ share the same index set. Two chords are crossed when two pairs of $H_{\text{SYK}}$ have some common elements in their sets of indices. See figure \ref{cd4} for all the possible chord diagrams of size $4$, this is enumerated by $\langle\text{tr} H_{\text{SYK}}^4\rangle_J$.  We refer to \cite{Berkooz} for the details of the connection between chord diagram enumeration and moments calculation of the SYK Hamiltonian.
\begin{figure}[H]
\begin{center}
\includegraphics[scale=.7]{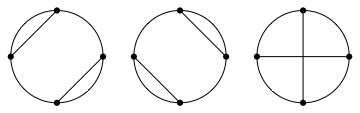}
\caption{$3$ possible chord diagrams of size $4$}
\label{cd4}
\end{center}
\end{figure}

Secondly, because we are in large $N$ limit, when we calculate the moments \eqref{moments3}, we only count products of consecutive $H_\text{SYK}$, because the extra source $\mathbb{\Lambda}$ only has one nonzero entry $\lambda_1$, the product of $\mathbb{\Lambda}$ sandwiched between any number of paired $H_\text{SYK}$ can be neglected. This means \eqref{moments3} can be written in terms of $\langle \text{tr} H_{SYK}^i \rangle^{k_i}_J$ and $\lambda_1^{p-\sum_{i}ik_i}$ with integers $k_i\in[0,\lfloor (p-1)/2\rfloor]$. Expanding \eqref{moments3} into its normally ordered form where all the $\langle\text{tr}H_{\text{SYK}}^{2i}\rangle_J$ are rearranged together following by all the $\lambda_1$. 
Replacing $\langle\text{tr}H_{\text{SYK}}^{2i}\rangle_J$ by $\text{RT}(i,q)$, we have
\begin{align}
\sum\text{RT}(1,q)^{k_1}\text{RT}(2,q)^{k_2}\cdots\text{RT}(l,q)^{k_l}\cdots \lambda_1^{p-2\sum_{i}ik_i}
\label{nf1}
\end{align}
where the summation is over all 
the combinations of products of $\text{RT}(i,q)$ and $\lambda_1$ with different orders that satisfy the total exponent equals $p$.
Besides, in \eqref{moments3}, we  subtract all the contributions only from SYK $\langle\text{tr}H_{\text{SYK}}^{2\lfloor p/2\rfloor}\rangle_J$. Accordingly, the exponent of $\lambda_1$ must be positive, then  $2\sum_i i k_i<p$.
Now, in order to make a connection to combinatorial enumeration, we need to do some adjustments to this product. We first multiplied each $\text{RT}(i,q)$ by $\lambda_1$; then the exponent in $\lambda_1^{p-2j}$ must be reduced to keep the product invariant 
\begin{align}
\sum(\lambda_1\text{RT}(1,q))^{k_1}(\lambda_1\text{RT}(2,q))^{k_2}\cdots(\lambda_1\text{RT}(l,q))^{k_l}\cdots \lambda_1^{p-2\sum_{i}ik_i-\sum_{i}k_i}.
\label{nf2}
\end{align}
Since each $\text{RT}(i,q)$ enumerates chord diagrams of size $i$ according to the number of crossings \cite{Riordan,Touchard}, the sum of 
all the possible normal forms above can be understood as an enumeration of a collection of chord diagrams with different sizes (including size $0$) basing on the total number of crossings, it is additionally invariant under cyclic permutation. 
Thus, this can be mapped to a counting problem of necklaces with each labeled bead attached to a chord diagram (one of the nodes in each chord diagram contacts directly to a labeled bead;  this node is called the root of the corresponding chord diagram).  
The multiplication of $\lambda_1$ in each parenthesis of \eqref{nf2} can be viewed as a marker for each occurrence of chord diagram. The rest $\lambda_1^{p-2\sum_{i}ik_i-\sum_{i}k_i}
$ marks the bead attached to a $0$-chord diagram (\.i.e. diagram with $0$ nodes and $\text{RT}(0,q)=1$). Such a necklace can have at most $2\lfloor(p-1)/2\rfloor$ nodes in the attached chord diagrams since we extract all the contributions that  only come from SYK Hamiltonian.
We need to label the beads, because the order of different $\text{RT}(i,q)$ contributed in \eqref{moments3} matters. 
An example of this kind of necklace is shown in Figure \ref{necklace1} with $p=10$ : bead $1$ is attached to a chord diagrams of $2$ nodes, bead $2$ is attached to a chord diagrams of $4$ nodes and $1$ crossing, bead $3$ is attached to $0$-chord diagram, thus marked by $\lambda_1^{10-(2+4)-2}=\lambda_1^2$ (for simplification, we don't show the $0$-chord diagram on the figure).
\begin{figure}[H]
\begin{center}
\includegraphics[scale=.7]{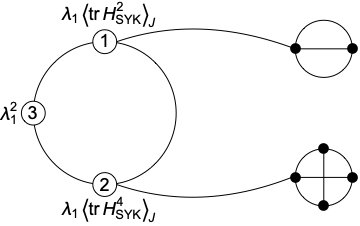}
\caption{A necklace of labeled beads attached to 3 chord diagrams: bead $1$ is attached to a $2$-chord diagram; bead $2$ is attached to a $4$-chord diagram and 
bead $3$ is attached to a $0$-chord diagram}.
\label{necklace1}
\end{center}
\end{figure}
More precisely, the connection between the calculation of \eqref{moments3} and the enumeration of the combinatorial structures described above is given by
\begin{align}
m_p=\sum_{\pi\in\mathcal{N}_{cd}}q^{\#\text{crossing}}
\label{momentsc}
\end{align}
where $\pi$ represents a specific labeled necklace of size $p$ composed of chord diagrams. 
Once we have this mapping, we can use the trick introduced in section \ref{sec:1} to get the enumeration of the right hand side of \eqref{momentsc}, or equivalently, the
subtracted moments \eqref{moments3}.
\subsection{Hanging chord diagrams in necklaces}
\label{subsec3.1}
A labeled necklace of chord diagrams is a necklace of beads on which chord diagrams are hung on and all the beads are labeled by integers. We denote by $\mathcal{N}_{cd}$ the set of such necklaces. The building blocks here are the chord diagrams, we denote it by $\mathcal{C}\mathcal{D}$. In addition, we use $\lambda_1$ to mark each of them. Thus, 
\begin{align}
\mathcal{N}_{cd}=\Theta\operatorname{CYC}(\lambda_1\mathcal{Z}\times\mathcal{C}\mathcal{D}) 
\label{opc}
\end{align}
$\mathcal{Z}$ represents the root of a chord diagram (the node connected to a bead). The cyclic structure comes
from the fact that a necklace is invariant by rotation. From \eqref{op_complex}, we get its generationg function
\begin{align}
N_{cd}(z):&=\sum_{p\geqslant 1}N_p z^p\nonumber\\
&=z\frac{d}{dz}\log\frac{1}{1-\lambda_1zR(z,q)}
\label{gf_opc}
\end{align}
where $N_p$ is the number of such necklaces with maximal $p-1$ nodes. $N_p$ enumerates the right hand side of \eqref{momentsc}. $R(z,q)$ is given by \eqref{gf_SYK},
which is the generating function of the chords diagrams enumeration.

We can reduce cycles to sequences
\begin{align}
\mathcal{A}=\operatorname{CYC}(\mathcal{B}) \quad\Rightarrow\quad \Theta \mathcal{A}=\mathcal{C} \operatorname{\times} \Theta \mathcal{B}, \quad\text{with}\quad \mathcal{C}=\operatorname{SEQ}(\mathcal{B}).
\label{eq_class}
\end{align}
This means a pointed cycle $\mathcal{A}$ of substructures $\mathcal{B}$ decomposes into the pointed component and the rest of the cycle; the directed cycle can then be opened at the place designated by
the pointing $\mathcal{B}$ and a sequence of $\mathcal{B}$. Therefore, combining \eqref{op_product}, \eqref{op_seq}, \eqref{op_cyc} 
, \eqref{op_pointing} and \eqref{eq_class}, we get 
\begin{align}
N_{cd}(z)=\left(1+\frac{z R^{\prime}(z,q)}{R(z,q)}\right)\frac{\lambda_1 zR(z,q)}{1-\lambda_1 zR(z,q)}.
\label{gf_opc2}
\end{align}
Since we know the mapping between moments calculation \eqref{moments3} and $N_p$, we prove our results \eqref{generationgF},   \eqref{generationgF2} and the equivalence between them.
Now by expanding \eqref{gf_opc2} or equivalently \eqref{gf_opc} and using \eqref{op_mix}, it follows that
\begin{align}
N_p:&=[z^p]N_{cd}[z]\nonumber\\
&=p\sum_{j=0}^{\lfloor\frac{p-1}{2}\rfloor}\sum_{\substack{k_1,k_2,\ldots,k_j \geq 0 \\ k_1+2k_2+\cdots + jk_j = j}} \lambda_1^{p-2j-\sum_{i}k_i}\frac{\binom{p-2j}{k_1,k_2,\ldots,k_j,p-2j-\sum_{i}k_i}}{p-2j} \prod_{i=1}^j (\lambda_1\text{RT}(i,q))^{k_i}
\label{Nn}
\end{align}
Therefore, by virtue of \eqref{momentsc}, we prove \eqref{moment1} through the mapping.
\section{Singularity analysis of generating function}
\label{sec:3}
In this section, we borrow some important notions from analytic combinatorics \cite{Flajolet2} which will be used later to identify the radius of convergence and the singular exponents of our composition structures. Our goal here is to investigate the effect of the input $\lambda_1$ on the spectrum of \eqref{hamiltonian} by using the singularity analysis of generating function \eqref{gf_opc2}.
This means we need to locate its singularities and find the corresponding singular exponents. In order to do that, we consider $\lambda_1$ as a variable. The idea here is that this auxiliary variable is regarded as inducing a deformation of the original univariate generating function (\.i.e. \eqref{gf_opc2} with $\lambda_1=\lambda^c_1$, which will be identified later). This deformation  can be analyzed by the corresponding asymptotic expansion of the generating function near its dominant singularities with a given $\lambda_1$. 
Therefore, \eqref{gf_opc2} becomes a bivariate generating function \cite[Chapter~\uppercase\expandafter{\romannumeral3}]{Flajolet2}, we rewrite it  as 
\begin{align}
N_{cd}(z,\lambda_1)&=\sum_{p,k}N_{p,k}z^p\lambda_1^{k}\nonumber\\
&=\left(1+\frac{z R^{\prime}(z,q)}{R(z,q)}\right)\frac{\lambda_1 zR(z,q)}{1-\lambda_1 zR(z,q)}.
\label{gf_opc3}
\end{align} 
where  $N_{p,k}$ is the number of necklaces in class $\mathcal{N}_{cd}$ of size $p$ having $k$ components of rooted chord  diagrams $\mathcal{Z}\times\mathcal{C}\mathcal{D}$ \eqref{opc}. We call $z$ and  $\lambda_1$ the primary and the secondary variable of $N_{cd}(z,\lambda_1)$, respectively. It is easy to see that
\begin{align}
N_p(\lambda_1):=\sum_{k}N_{p,k}\lambda_1^k\equiv[z^p]N_{cd}(z,\lambda_1)
\label{eqrgf}
\end{align}
\subsection{Singular expansions}\label{subsec:3.1}
We start by finding out where the singularities are located. As noted in \eqref{eq_class}, we can treat \eqref{gf_opc3} as a product of two parts.
Apparently, the sequence part $\mathcal{C}$ with the corresponding generating function 
\begin{align}
C(z,\lambda_1)=\frac{1}{(1-\lambda_1zR(z,q))}
\label{gfsq}
\end{align}
has singularities at $\rho_C(\lambda_1)$ satisfying 
\begin{align}
\rho_C(\lambda_1)R(\rho_C(\lambda_1),q)=1/\lambda_1.
\label{pteq}
\end{align}
The rest part of \eqref{gf_opc3} has constant singularities with given $q$, which does not change with $\lambda_1$. Therefore, we only need to focus on $\rho_C(\lambda_1)$ and its influence on \eqref{gfsq} to study the perturbations induced by $\lambda_1$.

Let's consider \eqref{gfsq} as a functional composition of two functions analytic at the origin that have non-negative coefficients
\begin{align}
C(z,\lambda_1)=f(\lambda_1g(z))\quad\text{with}\quad f(z)=\frac{1}{1-z}\quad\text{and}\quad
g(z)=zR(z,q).
\label{comp}
\end{align} 
Let $\rho_g$, $\rho_f$ be the radii of convergence of $g(z)$ and of $f(z)$, and define
\begin{align}
\tau_g=\lim_{z\rightarrow\rho^{-}_g}g(z)\quad\text{and}\quad \tau_f=\lim_{z\rightarrow\rho^{-}_f}f(z)
\end{align}
The singular expansion of \eqref{gfsq} depends on the values of $\rho_f$, $\tau_g$ and $\lambda_1$. $C(z,\lambda_1)$ is called critical if it satisfies $\lambda_1\tau_g=\rho_f\equiv1$. In this case, there is a confluence of singularities at $\rho_C=\rho_g$. Therefore, the collection of singular expansions is parameterized by $\lambda_1$ \footnote{we refer\cite[Chapter~\uppercase\expandafter{\romannumeral6}]{Flajolet2} for a more detailed presentation of singular expansions of  functional composition.}. In the following, we will show that there  exist discontinuities in the singular behavior of \eqref{gfsq} when the secondary parameter $\lambda_1$ traverses the special value $\lambda^c_1$.

By looking at the definition of $R(z,q)$ in \eqref{gf_SYK}, we find that $\rho_g=\sqrt{1-q}/2$ and the asymptotic of $g(z)$ around $\rho_g$ is
\begin{align}
&g(z)\sim \tau_g+c_g\sqrt{1-\frac{z}{\rho_g}}\quad\text{with $c_g<0$}\nonumber\\
&\text{and}\quad\tau_g= \sqrt{1-q}\sum_{k\geq0}(-1)^k q^{k(k+1)/2}
\label{singl}
\end{align} 
Since $q\in[0,1]$, $\tau_g$ is less than or equal to $1$. 
We distinguish $3$ cases of singular expansion of \eqref{gfsq} by the values of $\lambda_1$
\begin{itemize}
\item Obviously,   \eqref{gfsq} is critical at 
\begin{align}
\rho_C=\rho_g=\sqrt{1-q}/2 \quad\text{if} \quad \lambda_1=\lambda^{c}_1:=\frac{1}{\sqrt{1-q}\sum_{k\geq0}(-1)^kq^{k(k+1)/2}}
\label{critical}
\end{align}
we find the local singular expansion
\begin{align}
C(z,\lambda^c_1)\sim \left(\frac{-\lambda_1c_g}{\rho_f}\right)^{-1}\frac{1}{\sqrt{1-z/\rho_g}}\quad\text{where $\rho_f\equiv  1$}
\label{asymexp}
\end{align}
\item 
When $\lambda_1<\lambda^c_1$, we call it subcritical. In this case, $f(\lambda_1 z)$ is analytic at $\tau_g$. It is such that the perturbations induced by the $\lambda_1$ affect neither the location nor the nature of the basic singularity of \eqref{gfsq}. In this case, $\rho_C=\rho_g=\sqrt{1-q}/2$ is constant, which does not change with $\lambda_1$. On that basis expansions show that
\begin{align}
C(z,\lambda_1)&\sim f(\lambda_1\tau_g)-c\lambda_1f^{\prime}(\lambda_1\tau_g)(\lambda_1)\sqrt{1-\frac{z}{\rho_g}}\quad\text{with a constant $
c \in \mathbb{R}^{+}
$}
\end{align}
\item When $\lambda_1>\lambda^c_1$, we call it supercritical. In this case, $f(\lambda_1 z)$ is also analytic at $\tau_g$. Furthermore, there exists a value $r$ with $r<\rho_g$ such that $\lambda_1g(r)$ attains the value $\rho_f$, which triggers a singularity of $C(z,\lambda_1)$. In other words, $r\equiv\rho_C$ and $\rho_C$ here is no longer constant, it is the solution of \eqref{pteq}. Since around this point, $g(z)$ is analytic, the singular expansion of $C(z,\lambda_1)$ can be obtained by combining the singular expansion of $f(z)$ with the regular expansion of $g(z)$ at $r=\rho_C(\lambda_1)$
\begin{align}
C(z,\lambda_1)&\sim -\frac{1}{\lambda_1g^{\prime}(r)}\left(z-r\right)^{-1}\quad\text{where $r$ is a solution of \eqref{pteq}}
\label{superexp}
\end{align}
\end{itemize}
To conclude the three situations above, not only does the location of the singularity $\rho_C(\lambda_1)$ change with $\lambda_1$, but the corresponding singular exponent $\alpha_C(\lambda_1)$ also changes as $\lambda_1$ increases and crosses the critical value $\lambda^{c}_1$. They both experienced a non-analytic transition at $\lambda^{c}_1$.
We use a simple diagram to visualize changes in the singularity $\rho_C(\lambda_1)$ and the corresponding singular exponent $\alpha_C(\lambda_1)$, it is called phase-transition diagram \cite[Chapter~\uppercase\expandafter{\romannumeral9}]{Flajolet2}
$$
\begin{array}{cccc}\hline  \lambda_1\quad&\lambda^c_1-\epsilon \quad& \lambda^c_1 \quad& \lambda^c_1+\epsilon \\ \hline \rho_C\quad&\frac{\sqrt{1-q}}{2}\quad & \frac{\sqrt{1-q}}{2} \quad& r \\ \hline Z^{\alpha_C(\lambda_1)}\quad& Z^{1/2} & Z^{-1/2} & Z^{-1} \\ \hline\end{array} \quad Z:=\rho_C-z
$$
Thus, the location of the phase transition is at $\lambda^c_1$. In the following we use the results above to find the gap in  the spectrum of \eqref{hamiltonian}.
\subsection{Locating the gap} 
\label{subsec:3.2}
Here, we only need to consider the supercritical case since the gap in the spectrum of \eqref{hamiltonian} only appears when $\lambda_1>\lambda^c_1$. The singular expansion of \eqref{gf_opc3} in this case is a product of \eqref{superexp} and the singular expansion of $\lambda_1zg^{\prime}(z)$ around $z=r$. Because $g(z)$ is analytical at $r$, the singular expansion of \eqref{gf_opc3} around $z=r$ can be written as
\begin{align}
N_{cd}(z,\lambda_1)\sim  \frac{1}{1-z/r}
\label{sinexp}
\end{align}
Then we can use a simple identity
$$
\int_{-\infty}^{+\infty}\frac{\delta(s-a)}{1-z s}\text{d}s=\frac{1}{1-az}
$$ 
to identify the corresponding density of states $\rho(E)$ satisfiying 
$$
\int_{-\infty}^{+\infty}\rho(E)\frac{1}{1-z E}\text{d}E=\frac{1}{1-z/r}
$$
it gives 
$$
\rho(E)=\delta(E-1/r)
$$
Therefore, the gap is located at $1/r$ with $r$ being the solution of \eqref{pteq} and consequentially we get \eqref{ageq} from \eqref{pteq} via Stieltjes transform. 

As we demonstrated in this section, the perturbations induced by $\lambda_1$ display as the discontinuities in singular behavior of \eqref{gf_opc3}. It causes the phase transition in combinatorial counting and therefore causes the phase transition in the spectrum of \eqref{hamiltonian} via the specific mapping identified in section \ref{sec:3}. 
\section{Numerical comparison}
\label{sec:4}
We compare our analytic expressions with numerical results of  \eqref{hamiltonian} at finite $N$ and fixed $q=4$. Although all the  expressions obtained in this paper are under the double scaled limit \eqref{double-scaled},  they are still  good approximations to the exact results for fixed $q\ll N$. The reason for this is that SYK model itself has a very good agreement between the double scaled analytic expressions and its exact numerical results for finite fixed $q$, even for low values of $N$ \cite{Garcia1}. We won't discuss the calculation of the $p$-dependent corrections here. We refer to \cite{Jia,Berkooz} for more details about how to calculate analytically the difference between fixed $p$ case and the double scaled limit. In the following, we first test the correctness of our moments expressions \eqref{moment1} and \eqref{moment4}. Then we show numerical evidence of a phase transition that happens in spectral density when $\lambda_1$ passes its critical values $\lambda^c_1$. Finally, we compare  the numerical results of the single eignevalue which is split from the SYK spectrum when $\lambda_1>\lambda^c_1$
with our analytic prediction \eqref{ageq}.

We use Mathamatica to calculate the eigenvalues of \eqref{hamiltonian} with input $\lambda_1=3$ and $50$ samplings. Figure \eqref{numMom} shows the moments $m_p$ of $N=26$, $N=28$, $N=30$ and $N=32$ respectively.  
The numerical results are obtained by using these eigenvalues calculated by Mathematica and after spectral and ensemble averaging of the samples. Since we are comparing our analytical expressions with numerical results of the $p=4$ case, we need to replace all the $q$ in \eqref{moment1} and \eqref{moment4} by $\tilde{q}(4)$ where $\tilde{q}(p)$ is defined by \eqref{qbis}.
\begin{figure}[H]
\begin{center}
\includegraphics[scale=.43]{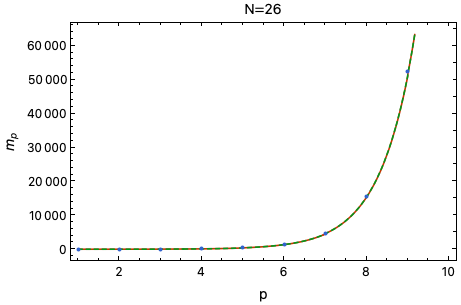}
\includegraphics[scale=.43]{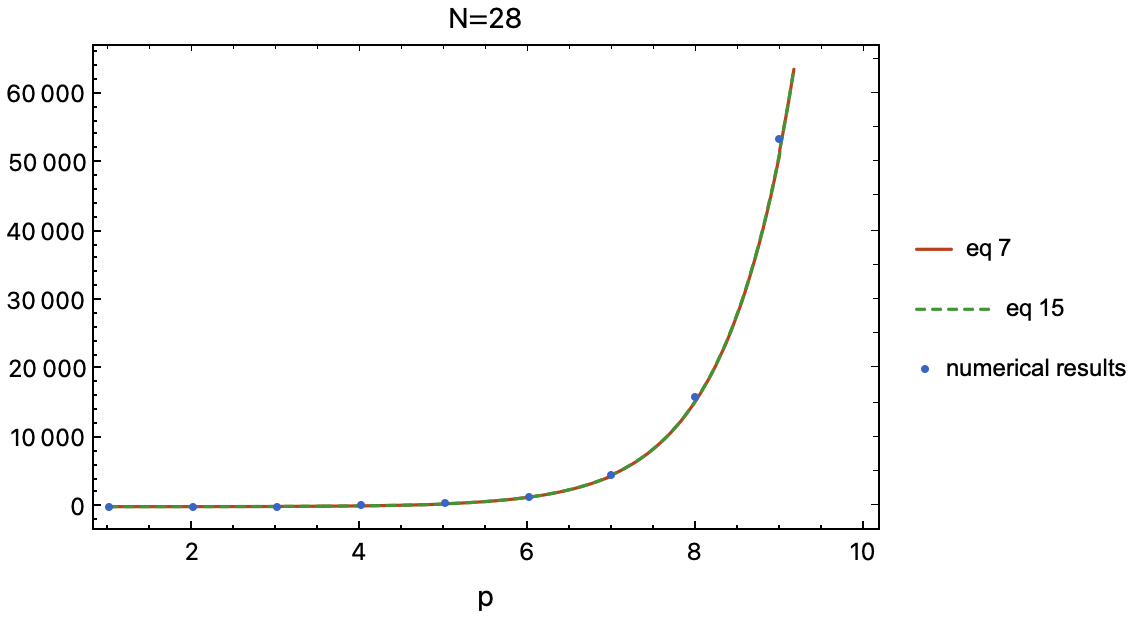}
\includegraphics[scale=.43]{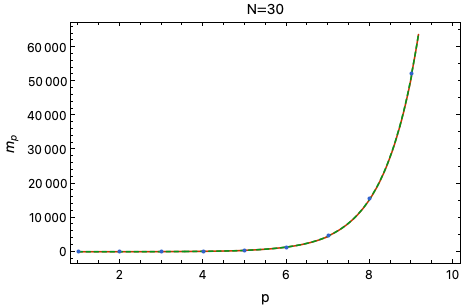}
\includegraphics[scale=.43]{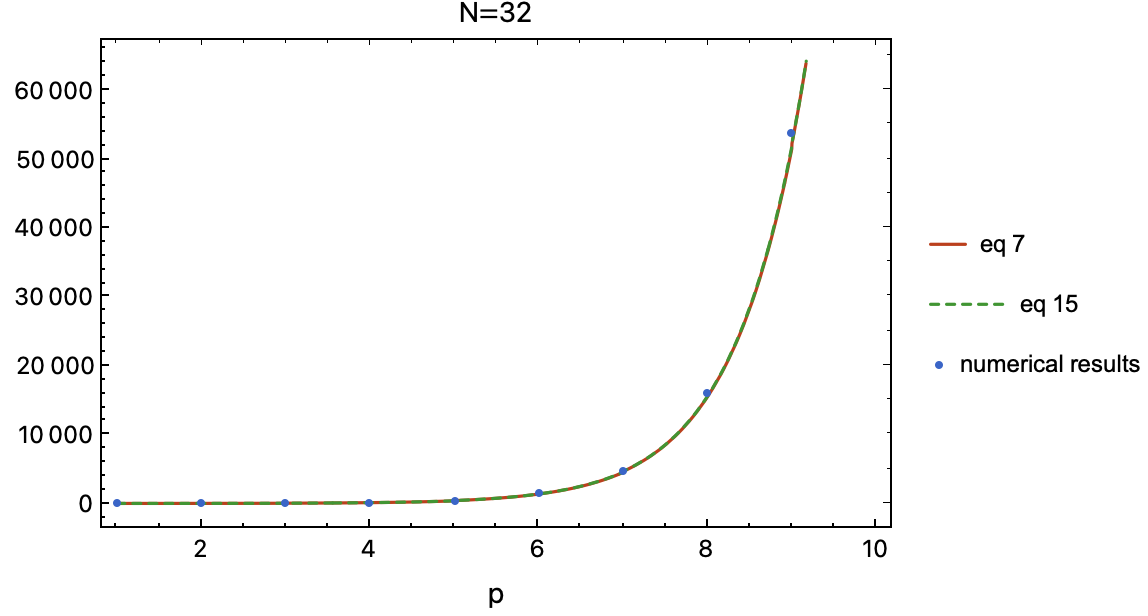}
\caption{Moments calculations comparison: the blue disk corresponds to the numerical results after spectral and ensemble average of $50$ samples. The red solid line is the analytical prediction \eqref{moment1}. The green dashed line is the analytical prediction \eqref{moment4}}
\label{numMom}
\end{center}
\end{figure}
As shown above, for the first few moments, we have excellent agreement between \eqref{moment1}, \eqref{moment4} and with numerical calculations.

Figure \ref{numdensity} below shows two examples ($N=24$ and $N=26$) of empirical spectral measure of $\eqref{hamiltonian}$ with two different values of $\lambda_1$. As  $\lambda^c_1$ defined in \eqref{critical}, for both $N=24$ and $N=26$ cases, $\lambda_1=1$ is smaller than $\lambda^c_1$ and $\lambda_1=3$ is larger than $\lambda^c_1$. One can see that as $\lambda_1$ passes its critical values,  there is a gap that appears in the spectrum. The main spectrum can be approximated by analytical expression of SYK's spectral density $\rho_{QH}(E)$ \eqref{densitySYK}. With given $\lambda_1>\lambda^c_1$, the single split eignevalue moves with $N$; hence it depends on $q$ as we predicted in the previous section.

In the following, we list the comparison between the numerical results and the analytic prediction of the averaged single split eigenvelue or $N = 24$, $N = 26$, $N = 28$, $N = 30$ and $N = 32$:
$$
\begin{tabular}{l|c|c|c}\hline
\diagbox{N}{}&
  $<E^{\text{numer}}_{\text{split}}>$ & $E^{\text{analy}}_{\text{split}}$ & $\sigma_{\text{split}}$\\ \hline
24 & 3.33851005    &  3.33824460& 0.0620565\\ \hline
26 & 3.34104622    & 3.33979434 & 0.0612582\\ \hline
28 & 3.34994743    & 3.34131759 & 0.0508209\\ \hline
30 & 3.34204363 & 3.34280205& 0.0508951\\ \hline
32 & 3.35162725  & 3.34424154 & 0.0599583\\ \hline
\end{tabular}
$$
where $\langle E^{\text{numer}}_{\text{split}}\rangle$ are the sample means of numerical results of the split eigenvalue. Each sample has exactly one split eigenvalue, we denote the single split eigenvalue in the $i$-th sample by $E^{\text{split}}_i$. Thus, $<E^{\text{numer}}_{\text{split}}>=\frac{1}{50}\sum_{i=1}^{50} E^{\text{split}}_i $.
Analytic prediction of this single split eigenvalue $E^{\text{analy}}_{\text{split}}$ is given by\eqref{ageq}. $\sigma_{\text{split}}$ in the table test the deviation of the sampling  $E^{\text{split}}_i$ away from the analytical prediction $E^{\text{analy}}_{\text{split}}$, defined as
$$
\sigma_{\text{split}}=\sqrt{\frac{1}{50}\sum_{i=1}^{50}(E^{\text{split}}_i-E^{\text{analy}}_{\text{split}})^2}
$$  

In this section we have checked that our analytic expressions are in good agreement with those calculated numerically from the exact diagonalization of \eqref{hamiltonian} with $p=4$ and finite $N$. It should be noticed that our analytical results  also work fine for other $p$ fixed cases such as $p=3$.
\begin{figure}[H]
\begin{center}
\includegraphics[scale=.5]{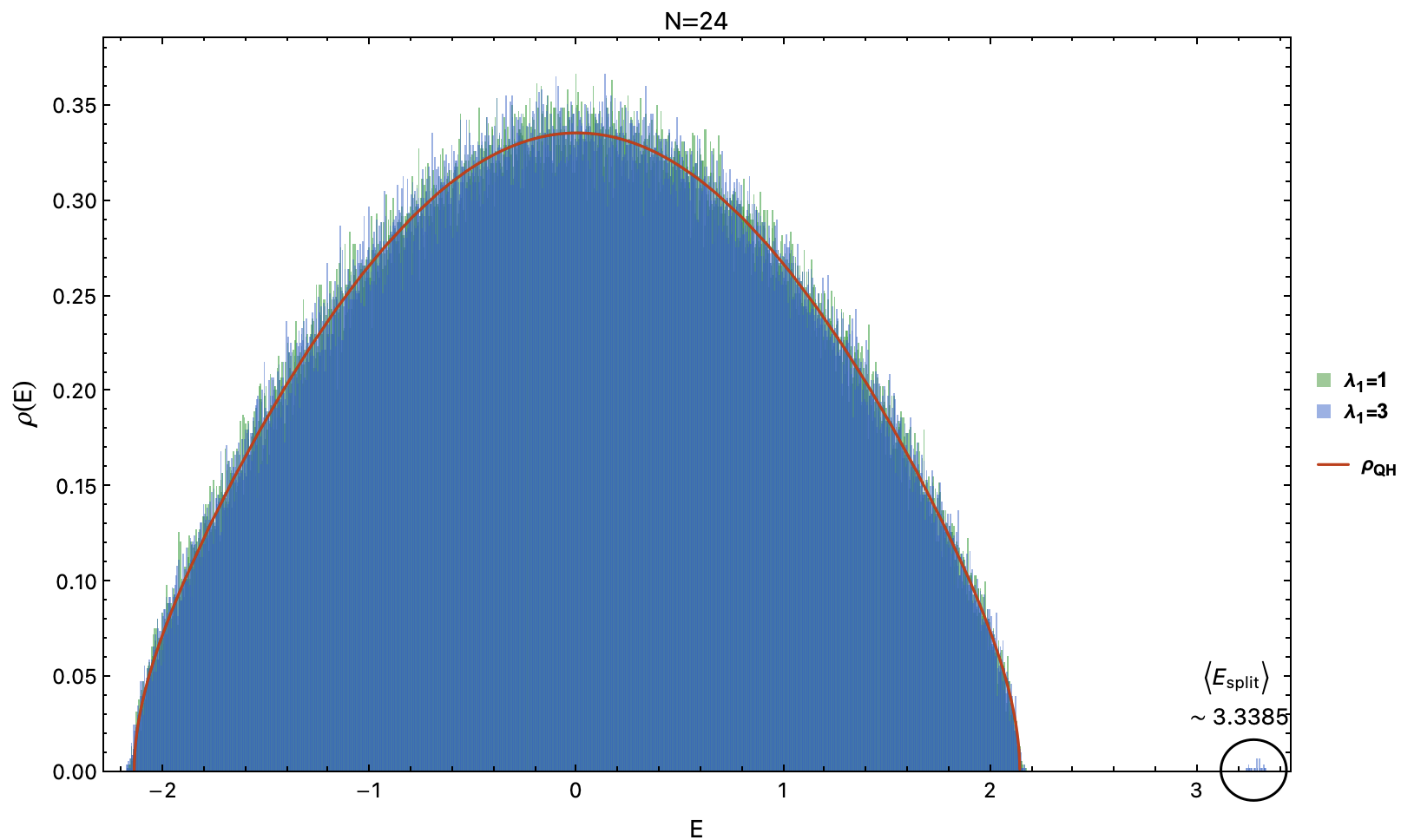}
\includegraphics[scale=.5]{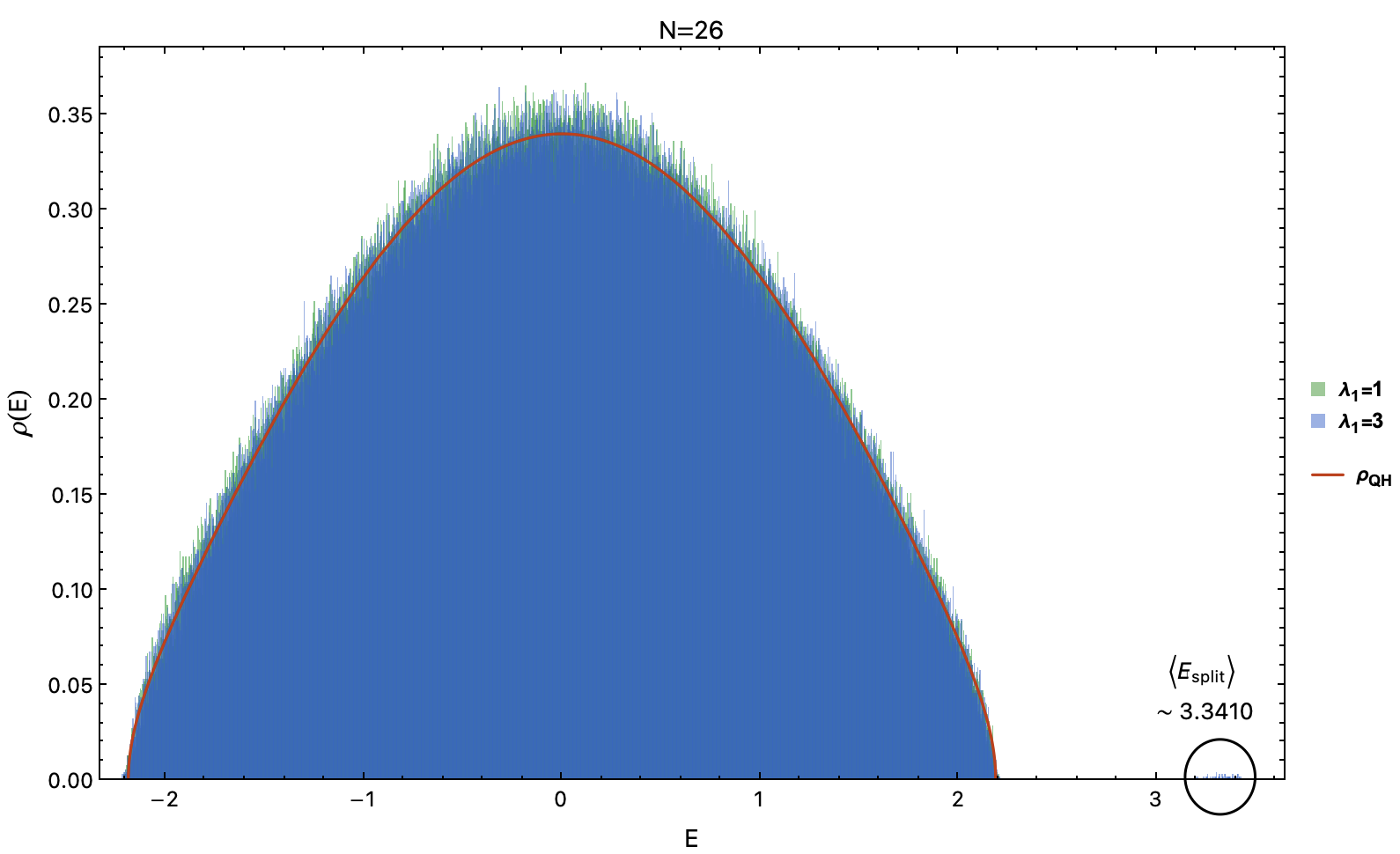}
\caption{Comparison of the numerical spectral density of \eqref{hamiltonian} with $p=4$, $\lambda_1=1$ (green shaded histogram) and $\lambda_1=2$ (blue shaded histogram) for $N = 24$ and $N = 26$, obtained by normalised histograms of eigenvalues of \eqref{hamiltonian} with sample size $50$, with the analytical prediction of the SYK spectral density $\rho_{QH}(E)$ (red smooth line). The sample mean of single split eigenvalue equals $3.3385$ for $N = 24$ (circled on the top figure) and equals $3.3410$ for $N = 26$ (circled on the bottom figure).}
\label{numdensity}
\end{center}
\end{figure}

\section{Conclusion}
\label{sec:5}
We have studied \eqref{hamiltonian} an SYK-like model with an extra part,  which is a simple diagonal matrix with only one non-zero entry $\lambda_1$. This model can be understood as an SYK model  ``poisoned'' by a rank $1$ perturbation.
Our analytic results not only are exact in the double scaled limit \eqref{double-scaled} but also are compatible with the $p$ fixed and $N$ finite cases. We have used the methods from analytic combinatorics to study the eigenvalue distribution of this model. Based on this, we have found that there is a phase transition in the secondary parameter $\lambda_1$ such that when $\lambda_1$ passes its critical value $\lambda^c_1$, a gap appears in the spectrum and a single eigenvalue splits outside the ensemble averaging SYK spectral density. We identified this eigenvalue by using singular analysis of a generating function which enumerates a specific class of combinatorial structures. The prime goal of our work is to better understand the precise effect of adding a deformation to the SYK Hamiltonian has on its eigenvalue distribution. Since \eqref{hamiltonian} only adds the smallest rank of perturbation, it can be used as a toy model to test the mapping between the phase transitions in physics and in combinatorial enumerations. Further research on more complicated higher rank perturbations can be built from here.
\section{Acknowledgments}
We would like to thank Antonio M. Garc{\'i}a-Garc{\'i}a for the interesting disscussion, useful suggestions and for a careful reading of the manuscript. We also acknowledge discussions with Hanteng Wang. More particularly,  we would like to thank
Stephan Wagner for communicating the derivations in the appendix.
\section*{Appendix}
We would like to have a closed form for the sum
$$S(p,j) = \sum_{\substack{k_1,k_2,\ldots,k_j \geq 0 \\ k_1+2k_2+\cdots + jk_j = j}} \binom{k_1+k_2+\cdots+k_j}{k_1,k_2,\ldots,k_j} \binom{p-2j}{k_1+k_2+\cdots+k_j} \prod_{i=1}^j C_i^{k_i},$$
where $C_i = \frac{1}{i+1} \binom{2i}{i}$ is a Catalan number. Let us first split the sum according to the value of $k_1+k_2+\cdots + k_j$:
\begin{align*}
s(p,j,r) &= \sum_{\substack{k_1,k_2,\ldots,k_j \geq 0 \\ k_1+2k_2+\cdots + jk_j = j \\ k_1+k_2+\cdots+k_j = r}} \binom{k_1+k_2+\cdots+k_j}{k_1,k_2,\ldots,k_j} \binom{p-2j}{k_1+k_2+\cdots+k_j} \prod_{i=1}^j C_i^{k_i} \\
&=
\sum_{\substack{k_1,k_2,\ldots,k_j \geq 0 \\ k_1+2k_2+\cdots + jk_j = j \\ k_1+k_2+\cdots+k_j = r}} \binom{r}{k_1,k_2,\ldots,k_j} \binom{p-2j}{r} \prod_{i=1}^j C_i^{k_i},
\end{align*}
so that
$$S(p,j) = \sum_{r=0}^{j} s(p,j,r).$$
We can rewrite $s(p,j,r)$ as follows:
$$s(p,j,r) = \frac{(p-2j)!}{(p-2j-r)!} \sum_{\substack{k_1,k_2,\ldots,k_j \geq 0 \\ k_1+2k_2+\cdots + jk_j = j \\ k_1+k_2+\cdots+k_j = r}} \prod_{i=1}^j \frac{C_i^{k_i}}{k_i!}.$$
The sum is exactly the coefficient of $x^j y^r$ in the expansion of
$$\prod_{i \geq 1} \sum_{k_i \geq 0} \frac{x^{i k_i} y^{k_i} C_i^{k_i}}{k_i!},$$
so we have
\begin{align*}
s(p,j,r) &= \frac{(p-2j)!}{(p-2j-r)!} [x^j y^r] \prod_{i \geq 1} \sum_{k_i \geq 0} \frac{x^{i k_i} y^{k_i} C_i^{k_i}}{k_i!} \\
&= \frac{(p-2j)!}{(p-2j-r)!} [x^j y^r] \prod_{i \geq 1} \exp (x^i y C_i) \\
&= \frac{(p-2j)!}{(p-2j-r)!} [x^j y^r] \exp \Big( y \sum_{i \geq 1} x^i C_i \Big) \\
&= \frac{(p-2j)!}{(p-2j-r)! r!} [x^j] \Big( \sum_{i \geq 1} x^i C_i \Big)^r \\
&= \binom{p-2j}{r} [x^j] \Big( \sum_{i \geq 1} x^i C_i \Big)^r.
\end{align*}
Thus
\begin{align*}
S(p,j) &= \sum_{r=0}^j \binom{p-2j}{r} [x^j] \Big( \sum_{i \geq 1} x^i C_i  \Big)^r \\
&= [x^j] \sum_{r=0}^{p-2j} \binom{p-2j}{r} \Big( \sum_{i \geq 1} x^i C_i  \Big)^r \\
&= [x^j] \Big( 1 + \sum_{i \geq 1} x^i C_i \Big)^{p-2j}.
\end{align*}
Note here that we are allowed to include terms with $r > j$ in the sum since the lowest power of $x$ in $\big( \sum_{i \geq 1} x^i C_i \big)^r$ is $x^r$, so that these terms do not actually contribute to the coefficient of $x^j$.

The generating function of the Catalan numbers is well known to be
$$F(x) = \sum_{i \geq 1} x^i C_i = \frac{1 - 2x - \sqrt{1-4x}}{2x},$$
it satisfies the implicit equation $F(x) = x (1+F(x))^2$. Therefore, we can apply the Lagrange-B\"urmann formula to obtain
\begin{align*}
S(p,j) &= [x^j] (1+F(x))^{p-2j} = \frac{1}{j} [t^{j-1}] (p-2j)(1+t)^{p-2j-1} (1+t)^{2j} \\
&= \frac{p-2j}{j} [t^{j-1}] (1+t)^{p-1} = \frac{p-2j}{j} \binom{p-1}{j-1} = \frac{p-2j}{p} \binom{p}{j}.
\end{align*}

\end{document}